\documentstyle[12pt]{article}
\newcommand{\LI}{\hbox to\hsize}
 \newcommand{\RLI}[1]{\LI{\hss#1}}

\newcommand{\PM}[1]%
{\mbox{$m_{\rm #1}$}} 
\newcommand{\BEQ}{\begin{equation}}
\newcommand{\EEQ}{\end{equation}}

\newcommand{\mr}[1]{\mbox{\rm #1}} 
\newcommand{\gdskd}{\vskip10mm
\begin{center}
G. Domokos and S. Kovesi--Domokos \footnote{E--MAIL: SKD@JHUP4.PHA.JHU.EDU}
\\
The Henry A. Rowland Department of Physics and Astronomy \\
The Johns Hopkins University \\
Baltimore, MD 21218
\end{center}
\vskip10mm }
\newcommand{\incircle}[1]{\mbox{{\hbox{$\bigcirc$}\kern-0.7em
\lower0.05ex\hbox{\mbox{{\scriptsize\rm #1}}}}}}
\newcommand{\eq}[1]{eq.~(\ref{#1})}
\newcommand{\ETC}{\mbox{\em etc.\/ }}
\newcommand{\VIZ}{\mbox{\em viz.\/ }}
\newcommand{\CF}{\mbox{\em cf.\/ }}
\newcommand{\IE}{\mbox{\em i.e. \/}}

\newcommand{\EG}{\mbox{\em e.g.\/ }}

\newcommand{\mathfig}[4]{%
\begin{figure}[#4] 
\vspace*{6in}
\centerline{\hbox to 14in 
 {\hskip5in {\special{eps:#2 x=14in}\hfil}}} 
\vspace*{-1.5in} 
\caption{#3}
\label{#1}
\end{figure}\vskip0.5in} 
\begin{document}
\RLI{JHU--TIPAC 94002}\vskip 1cm
\begin{center}
{\Large\bf SEMIANALYTIC CASCADE THEORY}\footnote{Invited talk
given at the third NESTOR workshop, Pylos 1993; to be published
in the Proceedings}
\end{center}
\gdskd
{\small We give a simple, semianalytic theory of the development
of the hadronic and neutrino/muon components of a cascade
induced by a primary which produces hadrons at the initial interaction.
The main purpose of the theory is to allow the user to obtain
quick, but reasonably reliable estimates of the longitudinal
properties of such cascades developing in a medium, such as a
stellar interior, the atmosphere and/or water. As an application,
we discuss the
possibility of discovering physics beyond the Standard Model by
means of neutrino telescopes. Some of those events may have
spectacular signatures in neutrino telescopes.}
\section{Introduction}

{\em ``Back of the envelope''} es\-ti\-mates play an im\-port\-ant
role in physics.
First, if one wants to decide the feasibility of an experiment, or
the observability of a phenomenon a theorist just discovered,
typically, one does not want to spend hours of CPU time on a substantial
computer in order to learn the answer. (It is ``no'' in 90\% of the
cases, anyway\ldots) Second, even {\em after} an elaborate
computation has been performed, one usually wants to obtain an
intuitive understanding of the result; in part, in order to
learn how to proceed, in part, perhaps, in order to discover some
bug in the program the presence of which is not obvious otherwise.

Cascade theory, necessary in order to conduct feasibility studies
for neutrino telescopes or to understand the generation of
neutrinos inside the point sources in the sky, among other things,
is notoriously time consuming and unintuitive if it is to be
sufficiently accurate. It is desirable therefore to design some
simple ``back of the envelope'' cascade theory for the
purpose outlined above.  In the age before large scale computation
became feasible, cascade theory was full of --- often unreliable ---
approximations and clumsy numerical computations (cf. the
review of Nishimura\cite{nishimura}). This was rendered
largely obsolete by the arrival of
high speed computers. Nevertheless, one can learn some tricks from the
``classics'' (for  good reviews, see \EG Rossi's book, \cite{rossi}
or Nishimura's review article just cited)
 and invent a few new ones in order to reduce
the amount of numerical
computation necessary for obtaining quick estimates. The purpose of the
work reported here is a first step in this direction; a similar
approach to the problem was taken recently by Lipari~\cite{lipari}.
\section{Approximations}
We outline the approximations used in developing the theory. Most of them
can be improved upon at some cost in computing time. However, even in
its present form, the theory is suitable for, say, astrophysical calculations,
where uncertainties in the input data (\EG stellar structure, \ETC) are
substantially bigger than the errors introduced by the approximations.
\begin{itemize}
\item Throughout this work, we use a one dimensional cascade theory in
the diffusion approximation. Transverse development can be added on later
in the diffusion approximation with relative ease. However, the
present version is adequate for purposes of a first orientation on
ultra high energy (UHE) processes.
\item We assume the validity of Feynman scaling in the one particle
inclusive cross sections, \VIZ
\BEQ
E\frac{d\sigma}{dE} = \sigma F\left( \frac{E}{E_{in}}\right),
\EEQ
where $E$ and $E_{in}$ stand for the observed and initial
particle energies, respectively and $\sigma $ stands for the total
inelastic cross section.

In practice, a two parameter fit to $F\left( z\right) $ \VIZ

\BEQ
F\left( z \right) = A \left( 1-z\right) ^{n}\Theta \left( 1-z\right)
\label{eq:fit}
\EEQ
gives a fair approximation
with $n \approx 3$, see, \EG \cite{collider}.
The distribution needs an infrared cutoff, in order to get a finite total
multiplicity. We choose a finite cutoff in $z$ in order to maintain Feynman
scaling and thus simplify the solution. In order to determine the
value of the infrared cutoff, somewhat arbitrarily, we chose a median
shower energy, $E_{L}\approx 5$ PeV, corresponding to a
CMS energy, $\sqrt{s}\approx 3$ TeV. At this energy, the {\em total}
average multiplicity is $\langle N \rangle \approx 30$, cf.\cite{databook}.
One knows that the infrared cutoff is of the order of
$\Lambda_{QCD}/\sqrt{s}$ which, with these numbers gives
$z_{0}\approx 10^{-4}$; in turn, that leads to $A\approx 4.1$.
(Most of our results are not very sensitive to the precise value
of $z_{0}$).

It is known that Feynman scaling is violated due to QCD loop effects;
in particular, the average multiplicity of a produced hadron,
\BEQ
\langle N\rangle = \int \frac{dz}{z}F\left( z \right)
\EEQ increases  with $E_{in}$. However, all such violations of
 Feynman scaling have a logarithmic dependence on the energy;
within the accuracy of the present calculations it is legitimate
to neglect them.  For the sake of consistency, one should then either
take all logarithmic dependences into account or none of them.

\item The total inelastic $\pi\mr{N}$ and $\mr{NN}$ cross sections
differ by about 30\% or so at high energies. Most of the
secondaries are pions, with pions of either charge being produced
in roughly equal numbers. There are few mesons produced containing s, c, b
(and t~?) quarks. The qualitative reason is that those
quarks have to be pulled out of a quark sea containing a small fraction
of heavy quarks. Likewise, baryon pair production is suppressed
due to the fact that a pair of three quarks has to be created coherently.
As a consequence, the total {\em high energy} baryon contents of
a high energy interaction is small: most of the baryons (with the
exception of a leading baryon in a $\mr{NN}$ interaction) come from
target fragmentation and they are of low energy in the LAB system.
These qualitative arguments are generally borne out by the experimental data,
wherever they are available,
see \EG \cite{databook}. In view of the above, it is justified to use
only one hadron distribution function as a first approximation. Instead of a
coupled system of integro--differential equations, we now have a
single equation describing the evolution of the cascade.
Due to the linearity of the eqations, whatever method
of solution is used for the single equation, it can be immediately
generalized to the more accurate, coupled system by a straightforward use
of matrix methods. However, the increase in computing time is not
negligible, since one has to invert matrices repeatedly.
\item Neutral pions drop out from the hadronic part of the
cascade development due to their
short lifetime at all energies of interest. Therefore, it is
justified to multiply the hadron distribution by a factor of
(2/3) in the cascade evolution: due to Pomeranchuk's theorem,
the $\pi^{\pm} \mr{N}$ cross
sections being very nearly equal. Conversely, in the development of
the electromagnetic component, the source of photons is fairly
represented by $1/3^d$ of the hadron distribution.
(The electromagnetic component
of a cascade in this approximation will be described elsewhere.)
\item Keeping the needs of neutrino telescopes in mind, it is
fair to say that a neutrino--rich shower is necessarily
a hadron--rich one:
the main source of UHE neutrinos is the process $\pi^{\pm}
\rightarrow \mu + \nu_{\mu}$, where we have not distinguished between
$\nu_{\mu}$ and $\overline{\nu}_{\mu}$. (The decay of muons is another
source of neutrinos; however, if one concentrates on the the UHE
part of the neutrino spectrum, $\mu$--decay neutrinos represent
only a small correction, see \cite{jourphysg}.)
Consequently, if the
initial interaction produces hadrons in substantial numbers,
either because the primary is a hadron or because of some ``new
physics'', such as the one conjectured in ref. \cite{strongneutrino}
or `t Hooft's B+L violating process, \CF A. Ringwald's
contribution to
these Proceedings, there will be a substantial number of neutrinos
present. Likewise, if one wants to determine, say, the UHE neutrino
spectrum emerging from an AGN or a binary system of stars, one can
concentrate on the accelerated hadrons (p, n, He, \ETC)
interacting with the target material, which often consists of
hadrons.

In turn, this circumstance drastically simplifies the description
of the evolution of the cascade. Due to the fact that
\BEQ
\sigma (\gamma ,hadron) \approx K \alpha \sigma (hadron, hadron),
\EEQ
($\alpha$ being the fine structure constant and $K$ a number of O(1)),
the feedback of the electromagnetic
component into the hadronic one represents a correction of about
1\% to the hadronic development. Neutrino--hadron interactions
have a much smaller cross section, typically scaled down by an
additional factor of $\left[ m_{\mr{h}}/m_{\mr{gauge}}\right] ^{2}$,
where the masses
involved are a typical hadron mass (say, 1 GeV) and a gauge boson
(W, Z) mass, of the order of 100 GeV.

Con\-sequent\-ly, the evo\-lu\-tion of the hadronic com\-po\-nent is, in
essence, an autonomous one. For the purposes of determining
the neutrino (and electron--photon) spectrum, the hadronic component
acts only as a source evolving autonomously.
\end{itemize}

\section{The Development of the Cascade}
We denote the differential distribution of hadrons by
$H\left( E, x\right)$, where $x$ stands for the depth
measured in units of the hadronic interaction mean free path. In this
manner, in the UHE region, one can adjust the units as new data become
available. With current data extrapolated to CMS energies of the
order of 50 to 100 TeV and averaging over $\pi \rm{N}$ and
$\rm{NN}$ cross sections, one gets $\lambda \approx 30 \rm{g/cm^2}$.
Using the approximations described in the previous Section,
we get the diffusion equation:
\BEQ
\frac{\partial H}{\partial x} = - \left(1 + \frac{2}{3}D
\right) H
+ \frac{2}{3}\int_{0}^{\infty} \frac{dE'}{E} H\left(E'\right)
F\left( \frac{E}{E'}\right) \Theta\left(E' - E\right).
\label{eq:diffusion}
\EEQ
Here the quantity $D$ stands for the loss of charged hadrons due to decay; its
expression is:
\BEQ
D\left( E, x\right) = \frac{\lambda m}{E \tau \rho \left( x \right)},
\label{eq:decay}
\EEQ
where $\rho \left( x\right)$ stands for the density of the medium
expressed as a function of $x$\footnote{Throughout this work
we use natural units, \IE $\hbar = c =1$.}. To a
good approximation, $m$ can be replaced by the pion mass and
$\tau$ by the charged pion lifetime, \IE $\tau \approx$ 7.8 meters.

Clearly, the term $D$
is quite small in the UHE region and at a first approach, it may
be neglected.

The expression of $\rho\left( x \right)$ depends on the model of
the medium used. For instance, assuming a radial incidence of the
primary, we have for some typical models:
\begin{enumerate}
\item Exponential atmosphere:
\BEQ
\rho \left(l\right) = \rho_{0} \exp \left[ - \frac{l}{h_{0}}\right],
\EEQ

\BEQ
\rho\left( x \right) = \frac{x \lambda}{h_{0}}.
\EEQ

\item Atmosphere with a power law for the density::
\BEQ
\rho = \rho_{1} \left[ \frac{l}{h_{1}} \right]^{-\kappa}, \quad
\left(\kappa > 1\right) ,
\EEQ

\BEQ
\rho\left( x \right) =
\rho_{1}
\left[ \frac{x\lambda}{\rho_{1}h_{1}}\right]^{\frac{\kappa}{\kappa-1}}.
\EEQ
\item For a (nearly) incompressible medium, like water, one may even
contemplate replacing $\rho$ by a constant, say, the value of the
density halfway between the entrance of the primary and the depth
of the detector.
\end{enumerate}
For a non--radial incidence, $l$ is to be
replaced by the slant depth: the changes in the formulae are
obvious and will not be exhibited here.
\section{Solutions}
\subsection{General techniques}
Neglecting the decay term in \eq{eq:diffusion}, --- an approximation valid
at the highest energies --- one can solve the equation by means of a Mellin
transformation in a standard fashion. The introduction of a new function,
$h\left( E, x\right)$ by means of the substitution
\[
H\left( E, x\right) = \exp\left( - x\right) h\left(E, x\right)
\]
reduces the equation to the form:
\BEQ
\frac{\partial h\left( E, x\right)}{\partial x } =
\int_{0}^{\infty} \frac{dE'}{E} F\left(\frac{E}{E'}\right)
\Theta \left( E' - E \right) h\left(E',x\right).
\label{eq:reduced}
\EEQ
This equation is to be solved with an initial condition suitable for the
physical problem at hand: we outline two types of such problems.
\begin{enumerate}
\item {\em Problems related to the development of a shower in an astrophysical
environment or a hadronic shower in the atmosphere.}

In that case, the primary spectrum can be well approximated by a power
spectrum for a substantial energy range. However, a power spectrum over
all energies is unphysical, since it has an infinite energy
contents\footnote{It is the consequence of this infinite ``energy
reservoir'' that a pure power spectrum passes through an absorber
without getting distorted.}.
(We also remark in passing that a pure power spectrum has no Mellin
transform either; thus, the popular factorized solution,
(see \EG \cite{rossi}, \cite{gaisser} or \cite{lipari}) {\em cannot} be
obtained
by using Mellin transforms, just by  substituting the {\em ad hoc}
Ansatz, $h = E^{-\alpha} f( x)$ into \eq{eq:reduced}.) One has to introduce
at least an ultraviolet cutoff, reflecting the fact that no physical
system is capable of producing particles of arbitrarily high energy.
Often, an infrared cutoff is also needed; however, in the present case,
this is not necessary: we are concentrating on the UHE part of the cascade.
Thus, energy--momentum conservation alone provides an effective
infrared cutoff.

In the following examples we use a spectrum of the form:
\BEQ
H\left( E, 0\right) = N \left( \frac{E}{E_{M}} \right)^{-\alpha}
\Theta \left( E_{M} - E \right)
\EEQ
\item {\em Problems connected with exploring some ``new physics''.}
In all models of physics beyond the Standard Model, one has to take into
account  the fact that the Standard Model is highly accurate up to
LEP and TEVATRON energies. As a consequence, it is commonly assumed
(for instance in \cite{strongneutrino}) that the onset of the ``new physics''
is sudden and that it can be characterized by a ``pseudothreshold'' in
$p_{T}$ and/or energy. Translated to the language of cascade development,
this means that all but the first interaction is described, in essence, by
the Standard Model. Consequently, the effect of the ``new physics'' can be
simulated by an initial condition imposed at the depth of the first
interaction,
\BEQ
x_{0} = \frac{\lambda_{new}}{\lambda},
\EEQ
where $\lambda_{new}$ is the mfp. corresponding to the new physics introduced.
(In practice, one may impose the initial condition at depth zero, but $x$
has to be replaced by $x-x_{0}$.)

{\em There is an important consequence of this picture, often ignored in the
literature.} Some authors have claimed that processes like `t~Hooft's instanton
induced B+L violating process and other processes, presently considered
exotic ones, can be recognized by searching for the occurrence of
high multiplicity muon bundles in neutrino detectors, roughly resembling a
heavy nucleus hitting the target (the atmosphere or water).
While this appears to be true for, say, $\nu$ and Pb induced showers
{\em started at the same depth}, in practice, $x_{0} \gg 1$ in any model
available: thus, care is needed in counting the multiplicity of muon bundles
in a neutrino telescope in order to find the onset of ``new physics''\ldots
\end{enumerate}

With this, the solution of \eq{eq:diffusion} is given by
\BEQ
H\left( E, x\right) = \mr{e}^{-x} \frac{1}{2\pi i} \int_{\cal{C}}
ds\, E^{-s}\tilde{H}
\left( s, 0 \right) \exp\left[ \tilde{F}\left(
s-1 \right) \, x\right],
\label{eq:solution}
\EEQ
where $\tilde{H}$ and $\tilde{F}$ stand for the Mellin
transforms of the initial condition and of $F$, respectively. The
contour $\cal{C}$ runs from $-i \infty$ to $+i\infty$ in the complex
$s$ plane, in the strip where both $\tilde{F}$ and
$\tilde{H}$ exist.

Equation (\ref{eq:solution}) can be a starting point of a numerical
evaluation of
the hadronic distribution. The initial conditions mentioned above
(and several variations on them) as well as the expression of $F$
are simple enough so that their Mellin transforms can be computed
analytically; yet, they are sufficiently accurate so as to give
a fair representation of the initial spectra and of the inclusive cross
sections. In general, however the integral in \eq{eq:solution}
cannot be evaluated in a closed form. Asymptotic methods used in the
distant past (saddle point, \ETC) are, in general, not
sufficiently accurate.

Another way, {\em very convenient from the point
of view of a numerical treatment}, is to solve the diffusion equation
by successive approximations. The convenient starting point for
this is \eq{eq:reduced}. Upon integrating both sides with respect
to $x$ and putting the equation in the form of a recursion relation,
we obtain:
\BEQ
h^{\left(n+1\right)}\left( E, x\right)= h\left(E, 0\right) +
\int_{0}^{x}\int_{E}^{\infty}\!dx' \frac{dE'}{E} F\left(
\frac{E}{E'}\right) h^{\left( n\right)}\left(E', x'\right).
\label{eq:successive}
\EEQ
Here $h^{\left( n\right)}$ stands for the $n^{th}$ iteration of the
solution, with
\[
h^{0} = h\left( E, 0\right).
\]

One notices that the integration over the variable $x'$ can be
performed in a closed form at every step of the iterative procedure.
 Moreover, with the simple representation of the inclusive
distribution suggested in this section and with the
simple initial spectra descibed above, the integral over the energy
can also be evaluated in a closed form, although the result is
somewhat clumsy. In this way, the iterative solution reduces to
the evaluation of a series. ({\em This, however,
may not be true for
more complicated representations of the initial inclusive
distribution and/or of the initial spectrum\/.})

A few remarks are in order here.
\begin{enumerate}
\item One knows that the iterative process just described is a convergent
one: \eq{eq:successive} is a Volterra equation which (due to the
compactness of its kernel) can always be solved by successive
approximations. (For purposes of proving this fact, one should
use $1/E'$ as an independent variable in \eq{eq:successive}).
As a consequence, if one uses the simple representations of the
inclusive cross section and initial spectra as described above,
the resulting series is a convergent one (in practice, it converges
quite rapidly).
\item The method of successive approximations described above
is, in essence, equivalent to the method of successive collisions
invented by Bhabha and Heitler in 1937, see \EG \cite{rossi}.
\item It is often advantageous to combine the numerical evaluation
of the cascade development with symbolic manipulation programs,
such as MATHEMATICA(C). In our experience, often a substantial amount
of computing time may be saved in this manner.
\end{enumerate}
\subsection{A Very Simple Solution}
If one is interested in the crudest type of estimates only, one
does not worry too much about the precise shape of the inclusive
distribution. In particular, one may just assume
that there is no leading particle present at all and the energy is
shared equally among all the secondaries, \IE
\BEQ
F\left( z \right) = \delta \left( z - \frac{1}{\langle N\rangle } \right).
\label{eq:delta}
\EEQ

The validity of this approximation has been discussed elsewhere,
see \cite{jourphysg}. We concluded in that reference that
the relative error
committed by using \eq{eq:delta} as opposed to a more realistic
form of the inclusive distribution is about a factor of $2$. Hence,
even this extremely crude approximation is adequate for the
purposes of qualitative estimates.
Using \eq{eq:delta}, one easily arrives at an explicit solution. Assuming
a primary power spectrum cut off at $E=E_{M}$, one gets:
\BEQ
H \left( E, x \right) = E^{-\alpha} \rm{e}^{-x}
\sum_{k=0}^{k_{max}}\frac{\left( x
\langle N\rangle ^{2 - \alpha}\right)^{k}}{k!}.
\label{eq:explicit}
\EEQ
The maximal value of $k$ is
given by:
\[
k_{max} =\left[ \frac{E_{M}}{E \ln \! \langle N\rangle }\right].
\]
Here $\left[ \cdots \right] $ stands for the integer part of a number.

Due to the sharp cutoff of the spectrum, $H\left( E, x\right) $
is a
function which has discontinuities. On could get rid of those discontinuities
by smoothing out in some way the $\Theta $ function occurring
in the spectrum: in that
case, however, one would be dealing with an infinite series.
Due to the crudeness of this approximation, we shall not discuss
this question any further. One notices that the approximate
solution written down here is a simple generalization of
``Heitler's caricature'' of an electromagnetic cascade,
\CF \cite{heitler}.

\subsection{Muons and Muon Neutrinos}

In the approximation used in the present work (UHE neutrinos and muons
only, no decaying muons, \CF the previous Section), there are
practically no
$\nu_{e}$ present: the branching ratio of $\pi ^{\pm}\rightarrow
e + \nu $ is of the order of $10^{-4}$, \CF \cite{databook}.
Further, the number of neutrinos and charged muons is equal
and determined by the autonomously evolving hadronic component:
\[
N\left( \nu + \overline{\nu} \right) = N\left( \mu ^{+} + \mu ^{-}\right)
\stackrel{def}{=} N.
\]

We get in an obvious manner:
\BEQ
N\left( E, x \right) = \frac{2}{3} \int_{0}^{x}\!dx'\, D\left( 2E, x'\right)
H\left( 2E, x' \right).
\label{eq:muons}
\EEQ

Thus, once the hadronic component is computed by using one of the
solutions described in the previous subsections, the number of UHE
muons and muon neutrinos is obtained by a quadrature. (It is
known of course that, either in the case of a pure power
spectrum, or going to lower energies in the present calculation,
electron neutrinos begin to appear. At energies substantially lower
than the ultraviolet cutoff,
the ratio of muon and electron neutrinos is about 2.)

Given the simplifications made in the course of the calculations,
one should check the accuracy and internal consistency of the results.

First, an internal consistency check. We stated earlier that the
iterative solution to the cascade equation should be convergent on
general grounds. In practice, however, the speed of the convergence
is an important factor from the computational point of view. For
this reason, we computed the hadronic flux and the atmospheric neutrino flux
generated by it
at a depth of 1000 ${\rm g/cm^{2}}$ and with a primary spectrum
$\propto E^{-\alpha}$ cut off at $E_{M}=10^{11}{\rm GeV}$.
For the sake of definiteness, we chose
$\alpha = 2.7$.

The iterative solution can be written in the form:
\BEQ
H(x,E) \propto E^{-\alpha}{\rm e}^{-x} \sum_{n=0}^{\infty}
\left(ax\right)^{n}f_{n}\left( E/E_{M}\right),
\label{eq:hadronic}
\EEQ
where the constant $a$ is determined in terms of fractional moments of
the fragmentation function. One easily verifies that
as $E_{M} \rightarrow \infty$, all $f_{n}\rightarrow 1$ and the
solution goes over into the one obtained with the factorization
assumption.

In order to exhibit the speed of the convergence, we chose $f_{0}=1$.
In Fig.~\ref{fig:fnloge} we exhibited the first few coefficients
$f_{n}(E/E_{M})$.

\mathfig{fig:fnloge}{c:/math/figures/fnloge.ps}{The first six
coefficients of the iterative solution.
The consecutive coefficients are plotted on a varying gray scale,
 $f_{1}$ being the darkest and widest line.}{t}

One sees that  the iterative solution is converging quite well:
the area under the consecutive coefficients $f_{n}(E/E_{M})$
decreases rapidly.

Next, we compute the hadron distribution and from it, the neutrino flux in
the manner described above. For the sake of simplicity, we used a
standard expression of the flux, ref.~\cite{gaisser} all the way to the
cutoff.
In Fig~\ref{fig:atmosnu}  we plot the integral spectrum of
neutrinos, multiplied by $E^{2.7}$.

\mathfig{fig:atmosnu}{c:/math/figures/atmosnu.ps}{The integral
spectrum of atmospheric neutrinos at a
depth 1000 ${\rm g/cm^{2}}$.}{t}

If the factorized
solution were valid everywhere, this quantity would be
approximately constant in the energy range we are considering.

We see that this is reasonably well satisfied up to $E/E_{M} \approx
10^{-3}$. In this region, our result agrees to within 30\% or so,
with Lipari's\cite{lipari}, and also by MC calculations. Given the
simplicity of the
model developed here, the agreement is satisfactory.

At still higher energies, the effect of the ultraviolet cutoff can be clearly
seen. In principle, one could infer the existence of such a cutoff from
atmospheric neutrino observations. The main obstacle to such an
observation is, of course, the scarceness of events.
\section{Looking for ``New Physics''}
In this Section, we illustrate the method developed in the
previous Sections by asking a simple question:

{\em Assuming that there exists some ``new physics'' beyond the
Standard Model, do we have any hope of detecting it in a neutrino
telescope\/?}

The answer, of course, depends on the cross section and energy of
the onset of
the new processes. We have in mind, in particular, either the
phenomenon conjectured in ref.\cite{strongneutrino} or the
multiple production of gauge Bosons (with or without B+L violation)
as discussed by A. Ringwald in these Proceedings.

There is a large amount of theoretical uncertainty in
the nature of the phenomena in which any physics beyond the Standard
Model would manifest itself. However, one can determine with
relative ease the reactions in which it is virtually hopeless to look
for manifestations of some post--Standard--Model physics, unless
the effects are unexpectedly dramatic. Any such argument is
based, in essence, on the unitarity of the S--matrix. Probability is
conserved, hence, {\em one should not look for the new physics
in reactions with a large number of open channels\/.} Unless
a firm theoretical prediction is available --- as it was in the case
of the discovery of the W and Z bosons in a hadronic machine --- it
is virtually impossible to find the proverbial needle in a haystack\ldots

This criterion immediately tells us that hadronic reactions are, in
all probability, unsuitable in a search for new physics: there
are just too many channels open, producing mostly mundane
physics: typically, relatively soft pions.
Likewise,  electron pair production in
$\gamma$ -- nucleus collisions dominates in a medium with
an average Z of the order of 3 or larger
(water, earth, the atmosphere). Hence, again, ``mundane physics''  suppresses
any appearence of new physics. (This question was discussed in some
detail in ref.~\cite{arkansas}. In a recent work, Morris and Ringwald,
ref.~\cite{morris} reached a similar conclusion by means of a
rather sophisticated Monte
Carlo simulation.)

What remains therefore, is the realm of neutrino induced reactions:
any process predicted by the Standard Model has a cross section
of the order of a few nanobarns at LAB. energies around an
EeV or so. By contrast, the conjectured new processes may
reach cross sections of the order of $10^{-2}$mb, \CF
ref.\cite{strongneutrino}, \cite{morris}. Therefore, they should be
observable in neutrino induced reactions better than anywhere else.

Neglecting all effects of nuclear structure, there
is a very simple relationship between a mfp ($\lambda$)
and the cross section
of a reaction (\CF \cite{jourphysg} and references discussed there):
\BEQ
\lambda [{\rm g/cm}^{2}] = \frac{1670}{\sigma [{\rm mb}]}.
\EEQ
(At high energies, this relationship should hold to an accuracy
better than about 20\% for medium heavy nuclei.)

Thus, a $\nu$--induced reaction of a cross section,
$\sigma \approx 5 \times 10^{-3} {\rm mb}$ and incidence at a zenith angle,
$\theta = 0$, would, in the mean, produce its first interaction
very close to a detector like NESTOR. (At incidences lower than
vertical, the range of cross sections one can explore below the
one at vertical incidence, depends
on the geology around the detector and it needs a more
detailed investigation. In general, however, one can explore several orders
of magnitude in the initial cross section by scanning at lower zenith angles.
For purposes of illustration therefore, we
assume vertical incidence and a cross section of the order of
magnitude just
quoted.)

Due to the theoretical uncertainties surrounding the ``new physics'', we
made two simplifying assumptions.
\begin{enumerate}
\item The onset of the ``new physics'' is sudden: in practice, we
approximate it by a $\Theta$ function in the energy, as in the articles
in ref.~\cite{strongneutrino}. As a consequence,
after the initial interaction, the resulting shower tends to evolve
 according to
the Standard Model. (At extremely high energies where this is not
the case, the
primary fluxes are expected to be very low.)
\item The rapidity distribution of the produced particles in the initial
interaction follows an ``equipartition law'', \VIZ
\BEQ
F(z) = \delta \left( z - \frac{1}{N_{1}}\right),
\EEQ
where $N_{1}$ stands for the average hadronic multiplicity
in the first interaction: this quantity parametrizes what
we call ``new physics'' in this model.
\end{enumerate}

Rather than compute the spectrum of the hadronic component for a
given primary neutrino spectrum, we investigated the profile of
single events, with a primary energy $E = 10^{16}$ GeV,
corresponding to about $\sqrt{s} \approx 5$ TeV in the neutrino--nucleon
CMS.

We computed the development of the hadronic cascade with two initial
multiplicities ($N_{1}=5$  and $N_{1}=20$).

Both cascades are
supposed to have a primary incident at zenith angle, $\theta =0$
and an initial cross section around the value just quoted.
The results are displayed in Fig.~\ref{bomban5} and Fig.~\ref{bomban20},
respectively.

\mathfig{bomban5}{c:/math/figures/bomban5.ps}{The profile the
hadronic component of an `anomalous'
shower  with $N_{1}=5$.}{t}
In both
Figures, we plotted the integral hadron spectrum as a function of
$x=t/\lambda$ for energies $E > 10$ GeV and $E > 100$ GeV, respectively.
(Obviously, the {\em higher} curves correspond to the {\em lower}
threshold energies in both Figures.)

In the Figures, $\lambda$ is the usual hadronic mfp, given the
assumption that after the first interaction, the cascade develops
according to the Standard Model.

\mathfig{bomban20}{c:/math/figures/bomban20.ps}{The profile
of the hadronic component of an `anomalous'
shower with $N_{1}=20$.}{t}

The interesting feature of these showers is that {\em charged hadrons}
 containing
light quarks (u,d,s), for all practical purposes, {\em do not decay}.
(Indeed,
the interaction mfp is about 80 cm, to be compared
with decay mfps substantially larger than 10 meters.) Similarly,
the radiation length is approximately 36 cm. Hence, both the hadronic and
electromagnetic cascades evolve within a few meters in
real space. The `anomalous'
showers are characterized by a very large energy deposition within the
span of a few meters. In particular, a  large amount of
Cherenkov light is emitted in such events.

Some  features are worth noticing.
\begin{itemize}
\item The overall profile of the
longitudinal evolution of a  shower is rather insensitive to the
initial multiplicity: hence, theoretical uncertainties are unlikely
to have a drastic effect on the discovery potential of a given neutrino
telescope.
\item The events are ``muon poor'' for the reason stated above. In practice,
the only source of muons is the production of particles containing
heavy quarks (c, b, t). However, the production of
heavy quarks is expected to be suppressed
due to the scarcity of the latter in the quark sea. (In the case of
multiple W--production, the source of muons is the leptonic decay mode
of the gauge bosons, which is of the order of 30\%.)
\item Overall, one concludes that an underwater neutrino telescope may be
the detector of choice for the observation of the events discussed here.
Due to the large {\em local} energy deposition, such anomalous events
may be observable even somewhat beyond the usual attenuation length
of Cherenkov light (of the order of 50m at NESTOR). This question needs
further study.
\end{itemize}

{\em The potential beauty of a search for new physics in observations
of the type just described is that the signature is a very robust
one\/}. Events
in which a large amount of energy is released near the neutrino
telescope  are hard to miss,
even though one does not understand all the details of such
events at present.

One event of an apparently large energy release and multiplicity
has been reported recently by the KAMIOKANDE collaboration,
ref.~\cite{kamiokande}. It is amusing to speculate about the possibility that
the ``unusual event'' as described by the authors
may be the manifestation of a phenomenon just discussed. Should that be the
case, one would have the first experimental evidence for the
existence of some physics
by means of which the ills of the Standard Model could be cured.

\section{Discussion}
We believe that the formalism presented in this paper is useful in order to
explore  new phenomena both in an astrophysical context
(point sources, AGN\ldots) and in a particle physics one.
Neutrino telescopes are potentially sensitive to the observation of
$\nu + `quark'$ reactions at energies beyond the reach of current
accelerator--based research. Hence, despite the erratic nature
of point sources and the uncertainties associated with sources
like AGN or gamma -- bursters, they may have an important
role to play in exploring  particle physics beyond the
$\sqrt{s}\approx 1$ TeV barrier.
\section{Acknowledgement}
This research has been supported in part by the U.S. Department of
Energy under Grant \# DE--FG--02--ER40211.

We thank Paolo Lipari, Christine Kourkoumelis, Al Mann, Kenzo Nakamura,
Leonidas Resvanis  and Atsuto Suzuki
for useful conversations during the Nestor Workshop.

We also thank our hosts, Leonidas Resvanis and his colleagues at the
University of Athens as well as the Honorable Mayor J. Vrettakos
and the City Council of
Pylos for hosting this productive and enjoyable meeting. \vskip5truemm

\end{document}